\RequirePackage{lineno}
\documentclass[aps,prl,twocolumn,superscriptaddress,showpacs,floatfix,letterpaper]{revtex4}
\usepackage{xcolor}
\usepackage{graphicx}
\usepackage{multirow}
\usepackage{amsfonts}
\usepackage{amsmath}
\usepackage{amssymb}
\usepackage{hyperref}
\usepackage{color}
\usepackage{float}
\usepackage{multirow}
\usepackage{lineno}
\usepackage{siunitx}
\usepackage{accents}
\usepackage{booktabs}

\usepackage[none]{hyphenat}
\usepackage{microtype}

\AtBeginDocument{%
  \setlength{\heavyrulewidth}{0.08em} 
  \setlength{\lightrulewidth}{0.05em} 
  \setlength{\cmidrulewidth}{0.03em}  
}

\newcommand{\beq}{\begin{eqnarray}}
\newcommand{\eeq}{\end{eqnarray}}

\def\SKV{50}  
\def\SKDV{32.5} 
\def\overburden{1,000}
\def\NODPMT{1,885} 

\def\rockR{4} 

\def\mipE{10} 
\def\mipQL{21} 
\def\bremE{100} 
\def\binL{1.6}    
\def\binM{500}   
\def\binH{5000}  
\def\binHH{90000}

\def\zcut{-0.1} 
\def\lcut{7}    
\def\Nupmu{3989}  
\def\Nday{4269} 
\def\QLa{0}
\def\QLb{27}
\def\QLc{30}
\def\QLd{35}
\def\QLe{40}

\def\DSTa{3989}
\def\DSTb{2010}
\def\DSTc{1130}
\def\DSTd{533}
\def\DSTe{316}

\def\preMCa{3367}
\def\preMCb{1673}
\def\preMCc{931}
\def\preMCd{460}
\def\preMCe{288}

\def\postMCa{$3938^{+462}_{-488}$}
\def\postMCb{$1962^{+254}_{-285}$}
\def\postMCc{$1092^{+153}_{-183}$}
\def\postMCd{$538^{+81}_{-107}$}
\def\postMCe{$336^{+52}_{-75}$}

\def\preXsecA{$0.48 \pm 0.08$}
\def\preXsecB{\dots}
\def\preFluxErrA{$0.00 \pm 0.04$}
\def\preFluxErrB{$0.00 \pm 0.09$}
\def\preAstroNorm{$1.70 \pm 0.20$}
\def\preAstroIndex{$2.60 \pm 0.50$}
\def\preKinErr{$0.00 \pm 0.05$}
\def\preBgRatio{$0.03 \pm 0.03$}
\def\preDetSyst{$0.00 \pm 0.05$}

\def\epreFluxErrA{4}
\def\epreFluxErrB{9}

\def\postXsecA{$0.53 \pm 0.04$}
\def\postXsecB{$0.51 \pm 0.11$}
\def\postFluxErrA{$0.01^{+0.04}_{-0.03}$}
\def\postFluxErrB{$-0.01 \pm 0.04$}
\def\postAstroNorm{$1.69 \pm 0.20$}
\def\postAstroIndex{$2.56^{+0.57}_{-0.40}$}
\def\postKinErr{$0.01 \pm 0.05$}
\def\postBgRatio{$0.03 \pm 0.02$}
\def\postDetSyst{$0.00^{+0.04}_{-0.03}$}

\begin{document}

\newcommand{\makespace}{\vspace{3 mm}}
\newcommand{\DP}{\displaystyle}

 \title{TeV-scale neutrino cross-section measurement using upward through-going muons in Super-Kamiokande}

\date{\today}

\iftrue
\newcommand{\AFFicrr}{\affiliation{Kamioka Observatory, Institute for Cosmic Ray Research, University of Tokyo, Kamioka, Gifu 506-1205, Japan}}
\newcommand{\AFFkashiwa}{\affiliation{Research Center for Cosmic Neutrinos, Institute for Cosmic Ray Research, University of Tokyo, Kashiwa, Chiba 277-8582, Japan}}
\newcommand{\AFFipmu}{\affiliation{Kavli Institute for the Physics and
Mathematics of the Universe (WPI), The University of Tokyo Institutes for Advanced Study,
University of Tokyo, Kashiwa, Chiba 277-8583, Japan }}
\newcommand{\AFFmad}{\affiliation{Department of Theoretical Physics, University Autonoma Madrid, 28049 Madrid, Spain}}
\newcommand{\AFFubc}{\affiliation{Department of Physics and Astronomy, University of British Columbia, Vancouver, BC, V6T1Z4, Canada}}
\newcommand{\AFFbu}{\affiliation{Department of Physics, Boston University, Boston, MA 02215, USA}}
\newcommand{\AFFuci}{\affiliation{Department of Physics and Astronomy, University of California, Irvine, Irvine, CA 92697-4575, USA }}
\newcommand{\AFFcsu}{\affiliation{Department of Physics, California State University, Dominguez Hills, Carson, CA 90747, USA}}
\newcommand{\AFFcnm}{\affiliation{Institute for Universe and Elementary Particles, Chonnam National University, Gwangju 61186, Korea}}
\newcommand{\AFFduke}{\affiliation{Department of Physics, Duke University, Durham NC 27708, USA}}
\newcommand{\AFFgifu}{\affiliation{Department of Physics, Gifu University, Gifu, Gifu 501-1193, Japan}}
\newcommand{\AFFgist}{\affiliation{GIST College, Gwangju Institute of Science and Technology, Gwangju 500-712, Korea}}
\newcommand{\AFFuh}{\affiliation{Department of Physics and Astronomy, University of Hawaii, Honolulu, HI 96822, USA}}
\newcommand{\AFFicl}{\affiliation{Department of Physics, Imperial College London , London, SW7 2AZ, United Kingdom }}
\newcommand{\AFFkek}{\affiliation{High Energy Accelerator Research Organization (KEK), Tsukuba, Ibaraki 305-0801, Japan }}
\newcommand{\AFFkobe}{\affiliation{Department of Physics, Kobe University, Kobe, Hyogo 657-8501, Japan}}
\newcommand{\AFFkyoto}{\affiliation{Department of Physics, Kyoto University, Kyoto, Kyoto 606-8502, Japan}}
\newcommand{\AFFliv}{\affiliation{Department of Physics, University of Liverpool, Liverpool, L69 7ZE, United Kingdom}}
\newcommand{\AFFmiyagi}{\affiliation{Department of Physics, Miyagi University of Education, Sendai, Miyagi 980-0845, Japan}}
\newcommand{\AFFnagoya}{\affiliation{Institute for Space-Earth Environmental Research, Nagoya University, Nagoya, Aichi 464-8602, Japan}}
\newcommand{\AFFkmi}{\affiliation{Kobayashi-Maskawa Institute for the Origin of Particles and the Universe, Nagoya University, Nagoya, Aichi 464-8602, Japan}}
\newcommand{\AFFpol}{\affiliation{National Centre For Nuclear Research, 02-093 Warsaw, Poland}}
\newcommand{\AFFsuny}{\affiliation{Department of Physics and Astronomy, State University of New York at Stony Brook, NY 11794-3800, USA}}
\newcommand{\AFFokayama}{\affiliation{Department of Physics, Okayama University, Okayama, Okayama 700-8530, Japan }}
\newcommand{\AFFosaka}{\affiliation{Department of Physics, Osaka University, Toyonaka, Osaka 560-0043, Japan}}
\newcommand{\AFFox}{\affiliation{Department of Physics, Oxford University, Oxford, OX1 3PU, United Kingdom}}
\newcommand{\AFFqmul}{\affiliation{School of Physics and Astronomy, Queen Mary University of London, London, E1 4NS, United Kingdom}}
\newcommand{\AFFregina}{\affiliation{Department of Physics, University of Regina, 3737 Wascana Parkway, Regina, SK, S4SOA2, Canada}}
\newcommand{\AFFseoul}{\affiliation{Department of Physics and Astronomy, Seoul National University, Seoul 151-742, Korea}}
\newcommand{\AFFsheff}{\affiliation{School of Mathematical and Physical Sciences, University of Sheffield, S3 7RH, Sheffield, United Kingdom}}
\newcommand{\AFFshizuokasc}{\affiliation{Department of Informatics in
Social Welfare, Shizuoka University of Welfare, Yaizu, Shizuoka, 425-8611, Japan}}
\newcommand{\AFFstfc}{\affiliation{STFC, Rutherford Appleton Laboratory, Harwell Oxford, and Daresbury Laboratory, Warrington, OX11 0QX, United Kingdom}}
\newcommand{\AFFskk}{\affiliation{Department of Physics, Sungkyunkwan University, Suwon 440-746, Korea}}
\newcommand{\AFFtodai}{\affiliation{Department of Physics, University of Tokyo, Bunkyo, Tokyo 113-0033, Japan }}
\newcommand{\AFFtit}{\affiliation{Department of Physics, Institute of Science Tokyo, Meguro, Tokyo 152-8551, Japan }}
\newcommand{\AFFtus}{\affiliation{Department of Physics and Astronomy, Faculty of Science and Technology, Tokyo University of Science, Noda, Chiba 278-8510, Japan }}
\newcommand{\AFFtriumf}{\affiliation{TRIUMF, 4004 Wesbrook Mall, Vancouver, BC, V6T2A3, Canada }}
\newcommand{\AFFtokai}{\affiliation{Department of Physics, Tokai University, Hiratsuka, Kanagawa 259-1292, Japan}}
\newcommand{\AFFtsinghua}{\affiliation{Department of Engineering Physics, Tsinghua University, Beijing, 100084, China}}
\newcommand{\AFFynu}{\affiliation{Department of Physics, Yokohama National University, Yokohama, Kanagawa, 240-8501, Japan}}
\newcommand{\AFFllr}{\affiliation{Ecole Polytechnique, IN2P3-CNRS, Laboratoire Leprince-Ringuet, F-91120 Palaiseau, France }}
\newcommand{\AFFbari}{\affiliation{ Dipartimento Interuniversitario di Fisica, INFN Sezione di Bari and Universit\`a e Politecnico di Bari, I-70125, Bari, Italy}}
\newcommand{\AFFnapoli}{\affiliation{Dipartimento di Fisica, INFN Sezione di Napoli and Universit\`a di Napoli, I-80126, Napoli, Italy}}
\newcommand{\AFFroma}{\affiliation{INFN Sezione di Roma and Universit\`a di Roma ``La Sapienza'', I-00185, Roma, Italy}}
\newcommand{\AFFpadova}{\affiliation{Dipartimento di Fisica, INFN Sezione di Padova and Universit\`a di Padova, I-35131, Padova, Italy}}
\newcommand{\AFFkeio}{\affiliation{Department of Physics, Keio University, Yokohama, Kanagawa, 223-8522, Japan}}
\newcommand{\AFFwinnipeg}{\affiliation{Department of Physics, University of Winnipeg, MB R3J 3L8, Canada }}
\newcommand{\AFFkcl}{\affiliation{Department of Physics, King's College London, London, WC2R 2LS, UK }}
\newcommand{\AFFwarwick}{\affiliation{Department of Physics, University of Warwick, Coventry, CV4 7AL, UK }}
\newcommand{\AFFral}{\affiliation{Rutherford Appleton Laboratory, Harwell, Oxford, OX11 0QX, UK }}
\newcommand{\AFFwu}{\affiliation{Faculty of Physics, University of Warsaw, Warsaw, 02-093, Poland }}
\newcommand{\AFFbcit}{\affiliation{Department of Physics, British Columbia Institute of Technology, Burnaby, BC, V5G 3H2, Canada }}
\newcommand{\AFFtohoku}{\affiliation{Department of Physics, Faculty of Science, Tohoku University, Sendai, Miyagi, 980-8578, Japan }}
\newcommand{\AFFicise}{\affiliation{Institute For Interdisciplinary Research in Science and Education, ICISE, Quy Nhon, 55121, Vietnam }}
\newcommand{\AFFilance}{\affiliation{ILANCE, CNRS - University of Tokyo International Research Laboratory, Kashiwa, Chiba 277-8582, Japan}}
\newcommand{\AFFibs}{\affiliation{Center for Underground Physics, Institute for Basic Science (IBS), Daejeon, 34126, Korea}}
\newcommand{\AFFglasgow}{\affiliation{School of Physics and Astronomy, University of Glasgow, Glasgow, Scotland, G12 8QQ, United Kingdom}}
\newcommand{\AFFoecu}{\affiliation{Media Communication Center, Osaka Electro-Communication University, Neyagawa, Osaka, 572-8530, Japan}}
\newcommand{\AFFminn}{\affiliation{School of Physics and Astronomy, University of Minnesota, Minneapolis, MN  55455, USA}}
\newcommand{\AFFsilesia}{\affiliation{August Che\l{}kowski Institute of Physics, University of Silesia in Katowice, 75 Pu\l{}ku Piechoty 1, 41-500 Chorz\'{o}w, Poland}}
\newcommand{\AFFtoyama}{\affiliation{Faculty of Science, University of Toyama, Toyama City, Toyama 930-8555, Japan}}
\newcommand{\AFFbmcc}{\affiliation{Science Department, Borough of Manhattan Community College / City University of New York, New York, New York, 1007, USA.}}
\newcommand{\AFFnumazu}{\affiliation{National Institute of Technology, Numazu College, Numazu, Shizuoka  410-8501, Japan}}

\AFFicrr
\AFFkashiwa
\AFFmad
\AFFbmcc
\AFFbu
\AFFbcit
\AFFuci
\AFFcsu
\AFFcnm
\AFFduke
\AFFllr
\AFFgifu
\AFFgist
\AFFglasgow
\AFFuh
\AFFibs
\AFFicise
\AFFicl
\AFFbari
\AFFnapoli
\AFFpadova
\AFFroma
\AFFilance
\AFFkeio
\AFFkek
\AFFkcl
\AFFkobe
\AFFkyoto
\AFFliv
\AFFminn
\AFFmiyagi
\AFFnagoya
\AFFkmi
\AFFpol
\AFFnumazu
\AFFsuny
\AFFokayama
\AFFoecu
\AFFox
\AFFral
\AFFseoul
\AFFsheff
\AFFshizuokasc
\AFFsilesia
\AFFstfc
\AFFskk
\AFFtohoku
\AFFtodai
\AFFipmu
\AFFtit
\AFFtus
\AFFtoyama
\AFFtriumf
\AFFtsinghua
\AFFwu
\AFFwarwick
\AFFwinnipeg
\AFFynu

\author{N.~Bhuiyan}
\AFFkcl
\author{K.~Abe}
\AFFicrr
\AFFipmu
\author{Y.~Asaoka}
\AFFicrr
\AFFipmu
\author{M.~Harada}
\AFFicrr
\author{Y.~Hayato}
\AFFicrr
\AFFipmu
\author{K.~Hiraide}
\AFFicrr
\AFFipmu
\author{T.~H.~Hung}
\AFFicrr
\author{K.~Ieki}
\author{M.~Ikeda}
\AFFicrr
\AFFipmu
\author{J.~Kameda}
\AFFicrr
\AFFipmu
% ...

\collaboration{The Super-Kamiokande Collaboration}
\thanks{\begin{widetext}
Corresponding authors: N.\ Bhuiyan (\href{mailto:bhuiyannahid@ihep.ac.cn}{bhuiyannahid@ihep.ac.cn}) and T.\ Katori (\href{mailto:teppei.katori@kcl.ac.uk}{teppei.katori@kcl.ac.uk}).%
\end{widetext}}
\noaffiliation

\fi

\begin{abstract}
Neutrinos provide a unique probe of both particle physics and the high-energy universe, traversing astronomical distances with minimal interaction. Their charged-current scattering cross section encodes fundamental information about weak interactions and nucleon structure across a vast energy range, yet measurements at TeV energies remain sparse. Here we report the first determination of the flux-averaged muon neutrino and anti-neutrino charged-current total cross section using high-energy atmospheric neutrinos observed in Super-Kamiokande. Using 3989 upward through-going muon events collected over 4269~days, together with a Bayesian fit to atmospheric flux and detector simulations, we measure the flux-averaged charged-current cross section in the 500-5000~GeV range to be $\sigma/E_\nu=(0.51\pm 0.11)\times 10^{-38}$~cm$^2$GeV$^{-1}$, with the highest precision to date in the TeV regime. Our results are consistent with accelerator-based measurements at lower energies and collider-based measurements at higher energies, bridging a critical gap between accelerator experiments and neutrino telescopes. This work demonstrates the capability of large underground detectors to perform precision cross-section measurements with atmospheric neutrinos, opening a new window for probing Standard Model physics and potential new physics searches at multi-TeV energies. 
\end{abstract}

\maketitle

\clearpage
\onecolumngrid
\clearpage
\twocolumngrid

\subsection*{Neutrino Interactions}

Neutrinos are unique probes of both fundamental interactions and the high-energy universe. Their weak coupling allows them to escape dense environments and traverse astronomical distances, but also makes their interaction cross sections challenging to measure precisely. Over the past decades, neutrino observations across a vast energy range~\cite{Formaggio:2012cpf} have established flavour oscillations~\cite{Super-Kamiokande:1998kpq,SNO:2002tuh,T2K:2024wfn,T2K:2025wet}, revealed aspects of nucleon structure~\cite{MINERvA:2023avz}, and opened a new window on extreme astrophysical sources~\cite{Kamiokande-II:1987idp,Bionta:1987qt,Alekseev:1987ej,IceCube:2018dnn,IceCube:2022der,IceCube:2023ame}. Yet this broad reach is not matched by equally continuous measurements of neutrino interaction cross sections. In particular, the charged-current (CC) cross section in the TeV regime remains comparatively less known, even though this is precisely the range in which neutrinos become powerful probes a broad class of new-physics scenarios~\cite{Schneider:2021wzs,DiBari:2016guw}.

Existing measurements define the broad outlines of this landscape, but not a continuous map across energy. Accelerator-based experiments have measured CC neutrino cross sections extensively up to a few hundred GeV~\cite{NuTeV:2005wsg}, while neutrino telescopes have constrained the cross section at substantially higher energies using atmospheric and astrophysical neutrinos~\cite{IceCube:2017roe,IceCube:2020rnc,IceCubeCollaborationSS:2025zgz}. This leaves an extended intermediate region for which direct measurements are still comparatively limited. Recent collider neutrino observations have provided the first direct measurements in part of this interval, marking an important step toward more continuous coverage across the TeV scale~\cite{FASER:2024hoe,FASER:2024ref}, while remaining tied to the energy reach of present-day man-made sources.

\begin{figure}[h!]
\includegraphics[width=\columnwidth]{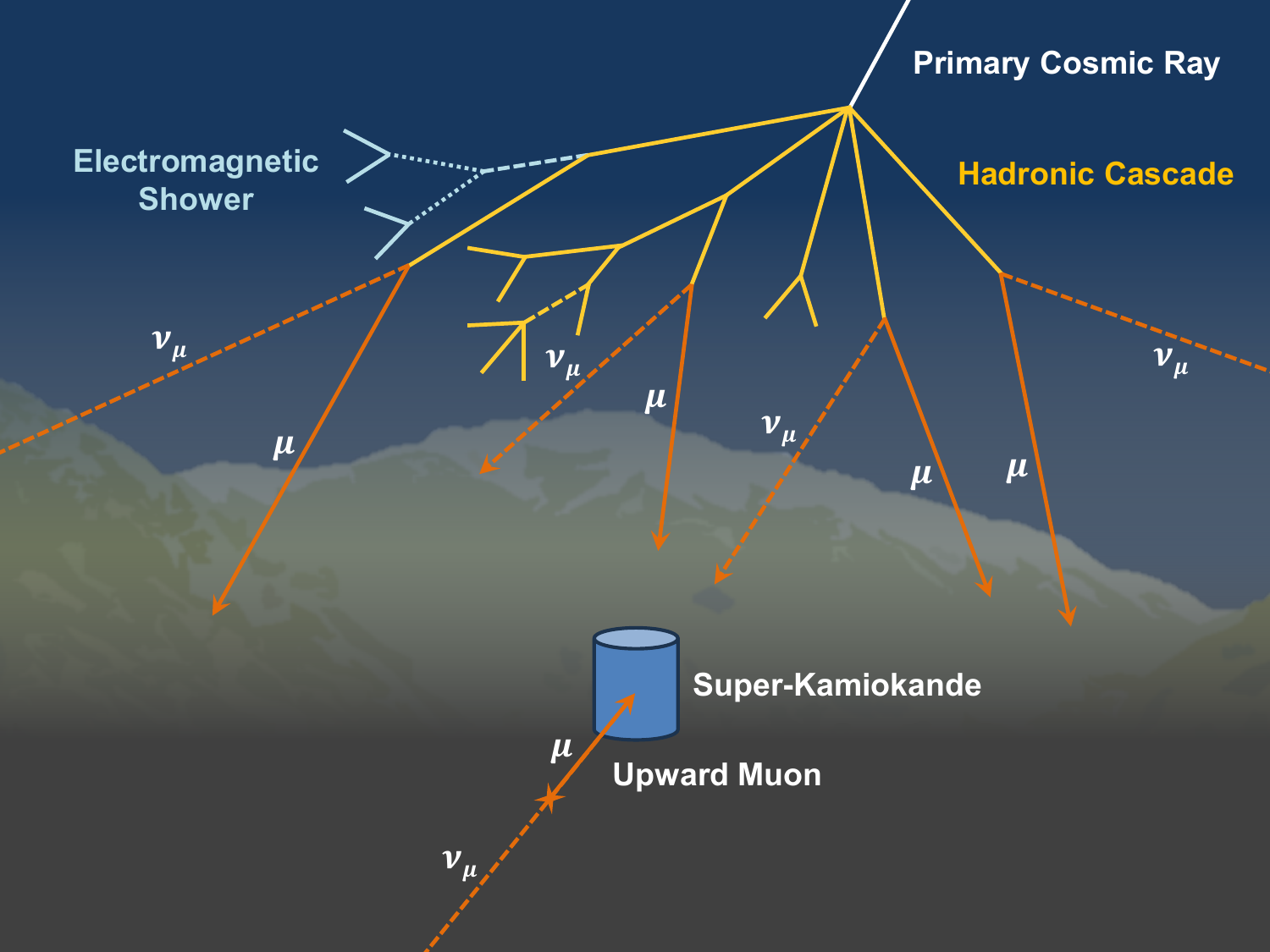}
\caption{Primary cosmic rays initiate hadronic cascades in the atmosphere, producing secondary mesons that decay into muons and muon neutrinos. Neutrinos generated in the southern hemisphere can traverse the Earth and undergo CC interactions in the rock surrounding Super-Kamiokande, producing upward muons that enter the detector.}
\label{fig:1}
\end{figure}

Atmospheric neutrinos offer a natural way to access this regime directly. Produced in cosmic-ray interactions in the Earth’s atmosphere, they provide a continuous flux extending into the multi-TeV range and beyond the energies readily accessible to controlled sources. Figure~\ref{fig:1} highlights the key geometry: atmospheric muon neutrinos can traverse the Earth and undergo CC interactions in the rock beneath large underground detectors, producing upward muons that enter and exit the detector volume. In this configuration, the Earth suppresses the otherwise overwhelming background of downward-going cosmic-ray muons, while the surrounding rock provides a large effective target mass for neutrino interactions. Upward through-going muons therefore provide a distinctive, background-suppressed signature of high-energy atmospheric neutrinos.

Here we present the first measurement of the flux-averaged muon-neutrino and anti-neutrino CC total cross section using high-energy atmospheric neutrinos observed in Super-Kamiokande~\cite{Super-Kamiokande:2002weg}. The detector is calibrated to measure neutrinos across a wide energy range, from low-energy solar neutrinos~\cite{Super-Kamiokande:2023jbt} to high-energy atmospheric neutrinos~\cite{Super-Kamiokande:2023ahc}, and has accumulated more than a decade of exposure, providing a uniquely large and well-characterised dataset for this study. We use the upward through-going muon sample~\cite{Super-Kamiokande:2023ahc}, thereby extending a mature underground neutrino programme into a direct cross-section measurement in the TeV regime.

\subsection*{The Super-Kamiokande Experiment}

Super-Kamiokande is a large water Cherenkov detector located approximately $\overburden$~m beneath Mt. Ikenoyama in Gifu, Japan~\cite{Super-Kamiokande:2002weg}. The detector consists of a cylindrical stainless-steel tank containing $\SKV$~kilotons of ultra-pure water, which is optically separated into an inner and an outer region. The inner detector holds $\SKDV$~kilotons of water and is instrumented with over 10,000 inward-facing 50~cm hemispherical photomultiplier tubes (PMTs) that record Cherenkov light emitted by charged particles. The inner detector is encased by the outer detector, which occupies the volume between the tank walls and the inner detector support structure. The outer detector is lined with reflective Tyvek sheets to improve light collection and is instrumented with $\NODPMT$ outward-facing 20~cm hemispherical PMTs. Each outer detector PMT is coupled to a wavelength-shifting plate to increase sensitivity to Cherenkov photons. Incident photons on a PMT generate photoelectrons (PEs), producing an electrical pulse. 
The charge $Q$ is an integrated electric pulse which is proportional to the number of PEs. The charge $Q$ and mean arrival time are then recorded by custom electronics~\cite{Nishino:2009zu}.

Through-going muons are topologically identified by clusters of PMT signals at the entry and exit points of the detector, as well as by the order in which the detector layers are passed. The recorded charge and timing information are used to reconstruct the muon’s track length $L$ and zenith angle $\theta_z$, where 
$\cos\theta_z=1$ is defined to point downward toward the Earth’s centre. 
Since downward-going muons from neutrino interactions are indistinguishable from cosmic-ray muons, only upward through-going muons are selected, yielding a clean sample of neutrino-induced events. The energies of the parent neutrinos extend into the TeV range, making this sample particularly sensitive to high-energy neutrino interactions. The detector does not distinguish between negative and positive muons; consequently, the sample contains contributions from both neutrino- and antineutrino-induced events, and the measured cross section corresponds to their flux-weighted average.

\subsection*{Data and Simulation}
This analysis uses $\Nday$~days of atmospheric neutrino data in Super-Kamiokande, selecting $\Nupmu$ events with an upward through-going muon-like topology, collected during the period from October 2008 to June 2022. To ensure that the entry and exit points are cleanly identifiable, only muons with reconstructed lengths $L > \lcut$~m are included. Near-horizon events, which are more susceptible to contamination from cosmic-ray muons, are further suppressed by requiring $\cos\theta_z < \zcut$. This pure muon sample contains approximately 3\% of backgrounds from three categories: the first arises from non-through-going muons that are misidentified and pass the selection; the second comes from through-going muons that are misidentified as upward-going; and the third consists of muons produced by processes other than 
muon neutrino CC interactions. The resulting sample is evaluated using these background estimates, together with models of the atmospheric neutrino flux, interaction kinematics, muon propagation, and detector response, to infer the neutrino interaction cross section from a comparison between the observed data and Monte Carlo simulations.

We use the 2014 Honda model~\cite{Honda:2015fha} for the atmospheric neutrino flux, smoothly connected to the Volkova spectrum~\cite{Volkova:1980sw} at 1~TeV. To evaluate flux-related uncertainties, we use the Daemon flux model~\cite{Yanez:2023lsy}, which provides a covariance treatment of 24 systematic parameters based on the primary cosmic ray and hadron production models. 
Since the Daemon model and the Honda-Volkova model are based on similar underlying assumptions, their predictions for the upward through-going muon rate differ by only about 1\% above 10~GeV. 
Our error estimate is therefore conservative, 
combining the covariance information from the Daemon model with the differences between the Honda-Volkova and Daemon flux models. This yields an atmospheric flux uncertainty of $\delta(\phi_0)=\epreFluxErrA$\% and $\delta(\phi_1)=\epreFluxErrB$\% corresponding to $\binL$--$\binM$~GeV and $\binM$--$\binH$~GeV, respectively; the rationale for these bounds is discussed in a later section.

{\it Conventional atmospheric flux} --- Using these atmospheric flux models, our simulation is primarily based on the contribution from muon neutrinos. The interactions of electron neutrinos resulting in detectable muons are negligible and therefore not considered. A small amount of muons from tau neutrinos is expected: at lower energies this arises from muon neutrino oscillations, but simulations show their contributions are very small and they are treated as background. At higher energies, tau neutrinos can also be produced via atmospheric charm decay~\cite{Bhattacharya:2015jpa}, but this lies outside the sensitivity of our analysis, and such a contribution has not yet been observed by neutrino telescopes~\cite{IceCube:2024fxo}, so it is considered negligible. 
Neutrino oscillations are incorporated directly in the Monte Carlo simulation through event-by-event weighting based on the atmospheric flux and neutrino oscillation probabilities~\cite{Super-Kamiokande:2023ahc}. For the upward through-going muon sample, the parent-neutrino energy spectrum is dominated by tens to hundreds of GeV, where the $\nu_\mu$ and $\bar\nu_\mu$ oscillation probabilities vary slowly after averaging over energy and baseline. 
Uncertainties in the oscillation parameters change the predicted event rate at only the $\mathcal{O}(1\%)$ level, making this contribution subdominant to other systematic effects and therefore negligible in the systematic uncertainty model.

{\it Astrophysical neutrino flux} --- The astrophysical contribution is modelled as a single power law of the form $\phi_\mathrm{astro} \cdot E^{-\gamma_\mathrm{astro}}$, where $\phi_\mathrm{astro}$ is the normalisation and $\gamma_\mathrm{astro}$ is the spectral index. Although such a component could contribute to the highest-energy neutrinos observed in Super-Kamiokande, its contribution to our upward-going sample is negligible ($\sim$1\%) based on measurements by IceCube~\cite{IceCube:2024fxo}. Accordingly, it is treated as background in this study with a large assigned uncertainty, $\phi_\mathrm{astro}=$\preAstroNorm~($10^{-18}$GeV$^{-1}$cm$^{-2}$s$^{-1}$sr$^{-1}$) and $\gamma_\mathrm{astro}=$ \preAstroIndex, with correlations, though it remains an interesting prospect for future analyses. 

{\it Neutrino interaction simulation} --- Neutrino interactions are simulated using NEUT \texttt{v5.6.4.1}~\cite{Hayato:2021heg}, a neutrino event generator that produces final-state particles given an input flux. At energies above $\sim$10~GeV, deep inelastic scattering (DIS) is the dominant interaction process and is the focus of this measurement. In NEUT, DIS interactions are simulated using the leading-order GRV98 parton distribution functions~\cite{Gluck:1998xa} with an additional correction~\cite{Bodek:2003wd}. Uncertainties in the DIS kinematics have a negligible effect on our analysis owing to the near-$2\pi$ coverage of both the neutrino flux and the detected muon phase space, which suppresses sensitivity to the differential cross-section shape. Consequently, our limited knowledge of the DIS differential cross-section has only a very small impact on the total cross-section measurement. Nevertheless, this effect is included as a systematic uncertainty through a shape variation of the inelasticity distribution. Specifically, we implement a global $\pm 5\%$ shift by reindexing the discretised inelasticity distribution, thereby shifting events towards higher or lower inelasticity while preserving the overall normalisation, following the guidance of Ref.~\cite{Weigel:2024gzh}.

{\it Detector simulation} --- The generated neutrino interactions are simulated within a spherical volume of radius $\rockR$~km surrounding the Super-Kamiokande detector. Beyond this distance, the probability for secondary particles to reach the detector becomes negligible due to geometric suppression and energy losses in the surrounding rock. The resulting final-state particles are then propagated into and through the detector using a GEANT3-based simulation~\cite{Brun:1987ma, Super-Kamiokande:2007uxr}, which accounts for the surrounding rock, water, and detector response. This simulated data sample is used to compare with the observed events and to extract the flux-averaged neutrino cross-section through our analysis framework.

\begin{figure}[h]
\includegraphics[width=\columnwidth]{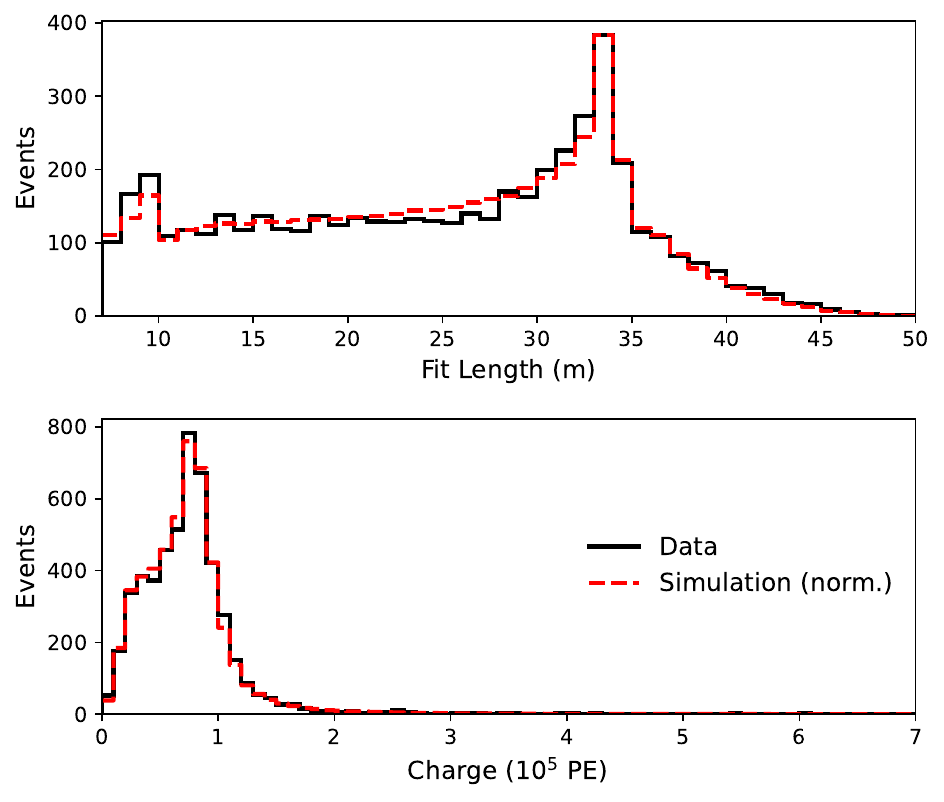}
\caption{Data-simulation comparisons of the fit track length $L$ and calibrated charge $Q$ distributions.}
\label{fig:fig2}
\end{figure}

Fig~\ref{fig:fig2} shows the data--simulation comparison of the fitted track length $L$ and charge $Q$ distributions. We consider systematic errors related to the detector, including the resolution of our measurements in $L$ (50~cm) and $Q$ (1\%), as well as the effective area of the detector, which together contribute a total uncertainty of approximately 5\%, referred to as detector effects.

\subsection*{Analysis}

\begin{figure}[H]
\includegraphics[width=\columnwidth]{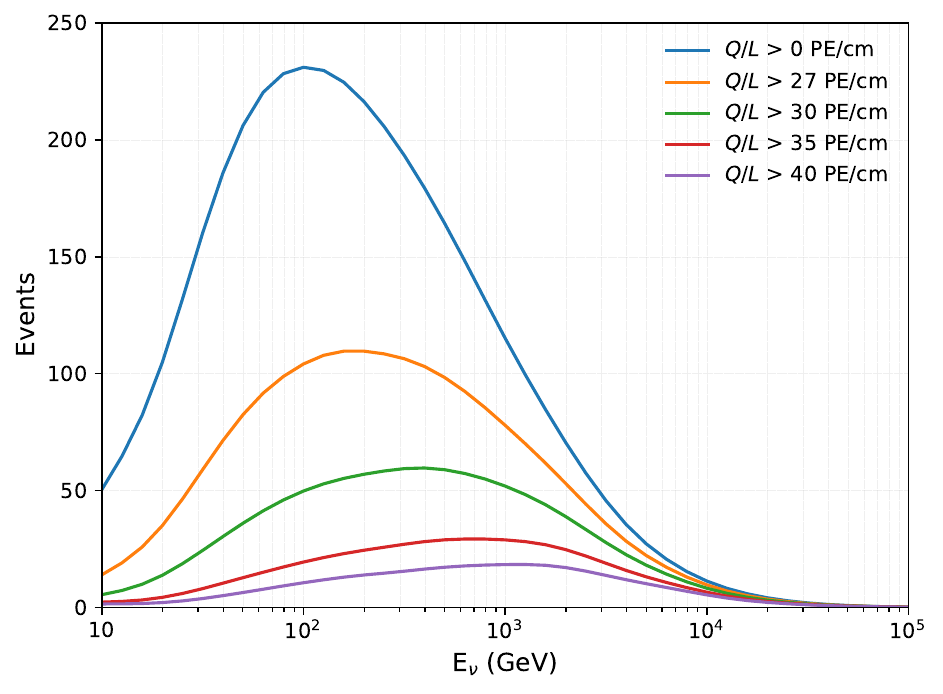}
\caption{Simulated upward through-going muon events in Super-Kamiokande. Events are grouped according to $Q/L$ (PE/cm) thresholds and smoothed.}
\label{fig:fig3}
\end{figure}

The key concept of this analysis is that the data and simulation can be subdivided into energy-sensitive subsamples using the relationship between $Q$ and $L$. Muons in Super-Kamiokande are typically minimum-ionising particles, for which the energy loss per unit length is nearly constant, around $\mipQL$~PE/cm, making $Q$ proportional to $L$. As a result, the detector can reliably measure muon energies only up to about $\mipE$~GeV. At higher energies ($\gtrsim\bremE$~GeV), radiative processes become increasingly important for the energy loss, causing $Q/L$ to exceed the minimum-ionising expectation, as illustrated in Fig.~\ref{fig:fig3}. Looser $Q/L$ selections retain more events but sample lower average energies, while tighter cuts increase the average energy at the cost of reduced statistics. Although muon energies cannot be precisely reconstructed on an event-by-event basis, the overall distribution of $Q/L$ provides the necessary information to infer the neutrino cross-section.

The observed event number $N$ can be schematically described 
by the convolution: 
\begin{equation}
\begin{split}
N = T \int R\cdot \left[\Phi_\nu\cdot P_\nu\cdot\sigma_{\nu}\cdot A_\mu 
+\Phi_{\bar\nu}\cdot P_{\bar\nu}\cdot\sigma_{\bar\nu}\cdot A_{\bar\mu} \right], 
\label{eq:event_rate_full}
\end{split}
\end{equation}
where $T$ is the exposure, $\Phi_\nu$ ($\Phi_{\bar\nu}$) is the atmospheric muon (anti-)neutrino flux, $P_\nu$ ($P_{\bar\nu}$) is the muon (anti-)neutrino oscillation survival probability, $\sigma_\nu$ ($\sigma_{\bar\nu}$) is the muon (anti-)neutrino CC cross section, $A_\mu$ ($A_{\bar\mu}$) is the negative (positive) muon attenuation during propagation, and $R$ is the detector response, encompassing both geometric acceptance and detection efficiency. Here, all of the parameters are functions of the energy and direction vectors of the respective particles. The observed events include contributions from both muon neutrinos producing negative muons and anti-neutrinos producing positive muons, and we treat the data as a flux-weighted average. To disentangle their combined contributions and extract the CC cross-section, we use a Bayesian approach with Markov Chain Monte Carlo (MCMC) simulations sampling the posterior with the No-U-Turn Sampler (NUTS)~\cite{Tool:Hoffman2014NUTS}, for a data--simulation fit.

In the fit framework, the simulation is divided into three energy regions: $[\binL,\binM)$~GeV, $[\binM,\binH)$~GeV, and $[\binH,\binHH]$~GeV. The lower bound, $\binL$~GeV, corresponds to the minimum muon energy capable of passing our selection criteria. The boundary between the first and second region, $\binM$~GeV, is chosen so that 
the first energy region can be guided using accelerator-based measurements, specifically $\sigma/E_\nu=(0.48 \pm 0.08) \times 10^{-38}$~cm$^2$GeV$^{-1}$ in our simulation. The boundary between the second and third region, $\binH$~GeV, is determined by evaluating the solution in 1~TeV increments; we find that above this point the results remain stable and no additional information is gained, indicating that higher-energy events contribute only minimally. Events above $\binH$~GeV are therefore treated as a low-statistics overflow. In this scheme, the prior for the cross-section in the first energy region, $\sigma_0$~($10^{-38}$~cm$^2$GeV$^{-1}$), is informed by existing accelerator-based measurements, while the second region, $\sigma_1$~($10^{-38}$~cm$^2$GeV$^{-1}$), is left unconstrained and sampled with a uniform prior. The third region accounts for excess high-energy events and is treated as an intrinsic uncertainty, and therefore not reported separately. This segmentation allows the fit to incorporate prior knowledge where available while remaining data-driven in less constrained energy ranges.

The fit is performed by achieving agreement in event numbers between data and simulation across multiple $Q/L$ regions, specifically $Q/L > \QLa$~PE/cm, $Q/L > \QLb$~PE/cm, $Q/L > \QLc$~PE/cm, $Q/L > \QLd$~PE/cm, and $Q/L > \QLe$~PE/cm. These regions are overlapping rather than exclusive, with the fraction of higher-energy muons increasing in the higher $Q/L$ thresholds. The correlations between these regions provide constraints on flux systematic uncertainties, particularly the overall normalisation. The fit uses the parameters listed in Tab.~\ref{tab:1}, and further details of the correlations are provided in the Supplemental Materials. The resulting posterior distributions of the fit parameters enable the extraction the flux-averaged CC cross-section as a function of neutrino energy.

\subsection*{Results}

\begin{figure*}[t]
\centering
\includegraphics[width=0.9\textwidth]{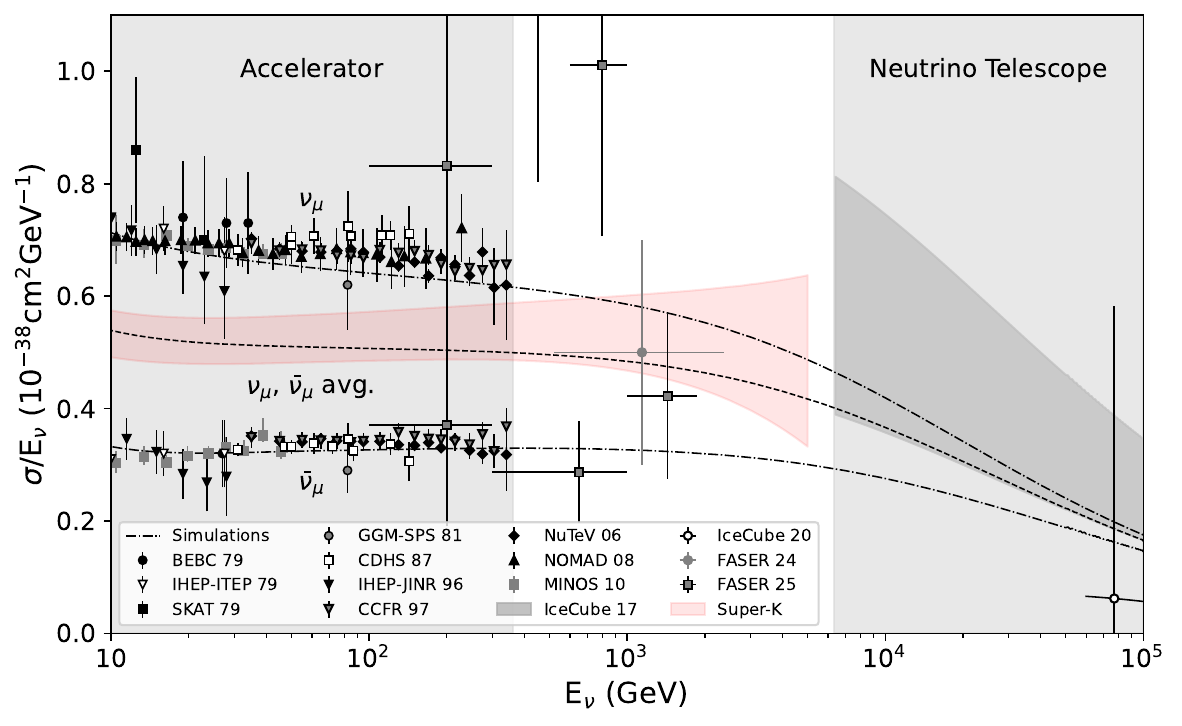} 
\caption{Measured CC muon neutrino and anti-neutrino flux-averaged cross section from the Super-Kamiokande upward through-going muon data. The red band shows the forward-folded result within the energy range 10--5000~GeV, with uncertainties propagated from the full inference. The dash-dotted lines show the simulated $\nu_\mu$ and $\bar{\nu}_\mu$ cross-section predictions used in this analysis~\cite{Hayato:2021heg}. The central dashed line shows their flux-weighted average, where the weighting is given by the atmospheric neutrino flux ratio in the Honda--Volkova model~\cite{Honda:2015fha,Volkova:1980sw}. The figure also includes external data from various experiments~\cite{MINOS:2009ugl,NOMAD:2007krq,Colley:1979rt,Berge:1987zw,Seligman:1997fe,GargamelleSPS:1981hpd,Mukhin:1979bd,Anikeev:1995dj,NuTeV:2005wsg,Baranov:1978sx,FASER:2024hoe,FASER:2024ref,IceCube:2017roe,IceCube:2020rnc}.
}
\label{fig:4}
\end{figure*}

The extracted cross-sections account for the uncertainties captured in the fit, including atmospheric and astrophysical fluxes, DIS kinematics, background variations, and detector effects. The prior and posterior values of these parameters, including their uncertainties, are summarised in Tab.~\ref{tab:1}. The flux-averaged muon neutrino and anti-neutrino charged-current total cross section in the range $\binM - \binH$~GeV is measured to be $\sigma/E_\nu=($\postXsecB$)\times 10^{-38}$~cm$^2$GeV$^{-1}$. Using the fitted parameters, we forward-propagate the posterior distributions through the underlying model to produce a continuous, banded representation of the flux-averaged CC muon neutrino and anti-neutrino averaged cross section as a function of neutrino energy, shown in Fig.~\ref{fig:4}.

\begin{table}[H]
\centering
\caption{Summary of fit parameters with prior constraints and posterior best-fit values. All systematic uncertainties are modelled with normal distributions. The first cross-section bin is constrained using base accelerator data with a normal prior, while the second cross-section bin is left unconstrained and assigned a uniform prior.}
\vspace{4pt}
\begin{tabular}{c c c r}
\toprule
Parameter & Prior & Constraint & \multicolumn{1}{c}{\ \ Best-fit} \\
\midrule
$\sigma_0$                  & Normal  & \preXsecA      & \postXsecA \\
$\sigma_1$                  & Uniform & \preXsecB      & \postXsecB \\
$\delta(\phi_0)$              & Normal  & \preFluxErrA   & \postFluxErrA \\
$\delta(\phi_1)$              & Normal  & \preFluxErrB   & \postFluxErrB \\
$\phi_\mathrm{astro}$       & Normal  & \preAstroNorm  & \postAstroNorm \\
$\gamma_\mathrm{astro}$     & Normal  & \preAstroIndex & \postAstroIndex \\
DIS Kinematics              & Normal  & \preKinErr     & \postKinErr \\
Detector Effects            & Normal  & \preDetSyst    & \postDetSyst \\
Background                  & Normal  & \preBgRatio    & \postBgRatio \\
\bottomrule
\end{tabular}
\label{tab:1}
\end{table}


Although the current measurement is limited by statistics, it achieves the smallest uncertainties among existing data in the TeV-scale energy region. The results are consistent with accelerator-based neutrino cross-section measurements at lower energies and with collider-based measurements at higher energies. While there is no direct overlap with neutrino telescope-based cross-section data at the highest energies, our results are expected to be compatible. 
In summary, we performed the first CC cross-section measurement using the through-going muon samples in Super-Kamiokande, resulting in the most precise neutrino cross-section measurement in the TeV energy region to date. This demonstrates the high-precision high-energy atmospheric neutrino measurement capability of this detector, and it opens up new research directions, including new physics searches~\cite{Schneider:2021wzs}. The remaining gap between Super-Kamiokande data and neutrino telescope data in 5--10 TeV can be addressed by the upcoming Hyper-Kamiokande experiment~\cite{Hyper-Kamiokande:2025fci}: with over eight times the fiducial volume of Super-Kamiokande, Hyper-Kamiokande will not only provide higher-statistics measurements of these events but will also extend sensitivity to higher-energy atmospheric neutrinos due to the continuous energy spectrum. 

\subsection*{Acknowledgment} 
We gratefully acknowledge the cooperation of the Kamioka Mining and Smelting Company. The Super-Kamiokande experiment has been built and operated from funding by the Japanese Ministry of Education, Culture, Sports, Science and Technology; the U.S. Department of Energy; and the U.S. National Science Foundation. Some of us have been supported by funds from the National Research Foundation of Korea (NRF-2009-0083526, NRF-2022R1A5A1030700, NRF-2022R1A3B1078756, RS-2025-00514948) funded by the Ministry of Science, Information and Communication Technology (ICT); the Institute for Basic Science (IBS-R016-Y2); and the Ministry of Education (2018R1D1A1B07049158, 2021R1I1A1A01042256, RS-2024-00442775); the Japan Society for the Promotion of Science; the National Natural Science Foundation of China (Grants No. 12375100 and 12521007); the Spanish Ministry of Science, Universities and Innovation (grant PID2021-124050NB-C31); the Natural Sciences and Engineering Research Council (NSERC) of Canada; the Scinet and Digital Research of Alliance Canada; the National Science Centre (UMO-2018/30/E/ST2/00441 and UMO-2022/46/E/ST2/00336) and the Ministry of  Science and Higher Education (2023/WK/04), Poland; the Science and Technology Facilities Council (STFC) and Grid for Particle Physics (GridPP), UK; the European Union’s Horizon 2020 Research and Innovation Programme H2020-MSCA-RISE-2018 JENNIFER2 grant agreement no.822070, H2020-MSCA-RISE-2019 SK2HK grant agreement no. 872549; and European Union's Next Generation EU/PRTR  grant CA3/RSUE2021-00559; the National Institute for Nuclear Physics (INFN), Italy.

\bibliographystyle{apsrev}
\bibliography{upmuxs}

\appendix
\setcounter{figure}{0}
\setcounter{table}{0}
\renewcommand{\tablename}{SUPPL. TABLE}
\renewcommand{\thetable}{\arabic{table}}
\renewcommand{\figurename}{SUPPL. FIG.}
\renewcommand{\thefigure}{\arabic{figure}}

\section*{Supplemental materials}

\subsection*{Number of events with different $Q/L$ cuts}

Suppl.Tab.~\ref{tab:suptab1} summarises the data and MC event counts after each successive $Q/L$ threshold cut. The pre-fit and post-fit values reflect the predicted rates before and after applying the fitted parameter values. The improvement in agreement illustrates the effect of the cross-section scaling in reproducing the observed data.

\begin{table}[H]
\centering
\caption{Event counts after successive $Q/L$ cuts, comparing data to MC predictions before and after the fit. The post-fit values reflect the parameter adjustments from the forward-folding inference.}
\vspace{4pt}
\begin{tabular}{l c c c}
\toprule
Cut & $\mathrm{MC_{pre-fit}}$ &$\mathrm{MC_{post-fit}}$ & Data  \\
\midrule
$Q/L>\QLa$ & \preMCa & \postMCa & \DSTa \\
$Q/L>\QLb$ & \preMCb & \postMCb & \DSTb \\
$Q/L>\QLc$ & \preMCc & \postMCc & \DSTc \\
$Q/L>\QLd$ & \preMCd & \postMCd & \DSTd \\
$Q/L>\QLe$ & \preMCe & \postMCe & \DSTe \\
\bottomrule
\end{tabular}
\label{tab:suptab1}
\end{table}

\subsection*{Systematic error parameter correlations}

\begin{figure*}[t]
\centering
\includegraphics[width=\textwidth]{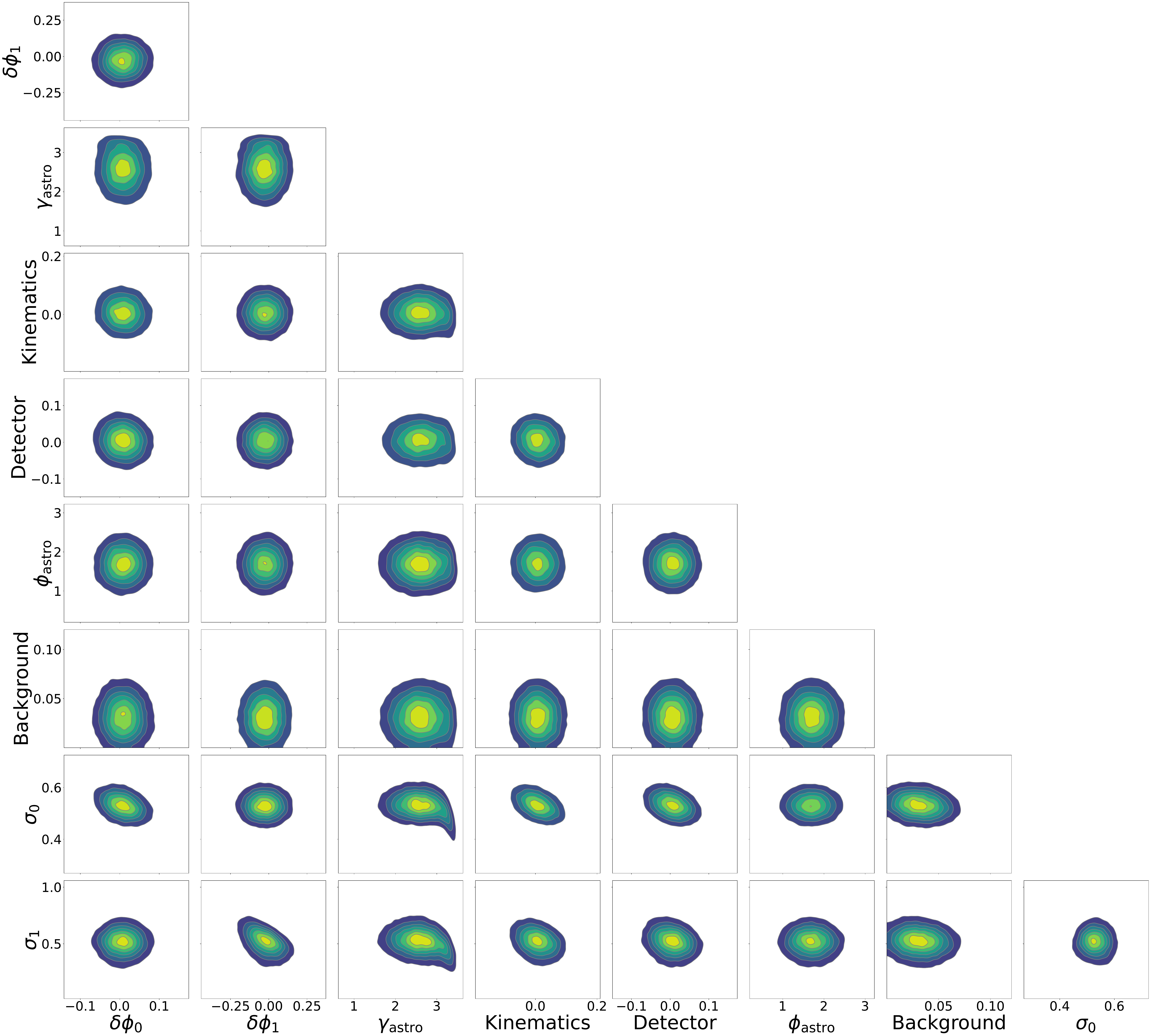}
\vspace{-2mm}
\caption{Corner plot showing the posterior distributions and parameter correlations from the forward-folding fit.}
\label{fig:supfig1}
\end{figure*}

The joint posterior distributions of all fit parameters are shown in Suppl.Fig.~\ref{fig:supfig1}. The two-dimensional contours illustrate the correlations between parameters, while the diagonal elements display the marginal distributions. This visualisation provides an overview of the parameter behaviour and their interdependencies in the fit, and we observe weak correlations across all parameters.

\end{document}